\documentclass[a4paper,12pt]{article}
\usepackage{graphicx,amssymb,bm,latexsym,color,epsf,cite,ulem,verbatim, physics, mathrsfs,lipsum, amsfonts,amsthm,stmaryrd, bm, hhline, float, empheq, geometry, tcolorbox, hyperref, etoolbox}
\tcbuselibrary{theorems}

\makeatletter
\@addtoreset{equation}{section}

\makeatother
\newcommand{\vs}[1]{\vspace{#1 mm}}
\newcommand*\di{\mathop{}\!\mathrm{d}}
\newcommand{\diff}{\mathscr{D}if\!f}
{
\addtolength{\skip\footins}{1.4pc plus 3pt}
\setlength{\footnotesep}{1.4pc}

\pagestyle{plain}

\textwidth 160mm
\textheight 230mm
\topmargin -10mm
\oddsidemargin 0mm
\begin{document}
	
	\newgeometry{top = 4cm,bottom = 4cm, left=2.4cm, right=2.4cm}

	\begin{center}
		{\LARGE Towards a Geometrization\vs{-2.5}\\of Renormalization
			 Group Histories\vs{+3}\\in Asymptotic Safety}
		
		\vs{10}
		
		{\large
			Renata Ferrero\footnote{e-mail address: rferrero@uni-mainz.de}
			and Martin Reuter\footnote{e-mail address: reutma00@uni-mainz.de}$^{}$
		} \\
		\vs{10}
	{\textit{Institute of Physics (THEP), University of Mainz,
				\\Staudingerweg 7, D-55128 Mainz, Germany}
		}
	\end{center}
\vs{5}

\setcounter{footnote}{0} 

\begin{abstract}
Considering the scale dependent effective spacetimes implied by the functional renormalization group in $d$-dimensional Quantum Einstein Gravity, we discuss the representation of entire evolution histories by means of a single, ($d+1$)-dimensional manifold furnished with a fixed (pseudo-) Riemannian structure. This ``scale-space-time" carries a natural foliation whose leaves are the ordinary spacetimes seen at a given resolution. We propose a universal form of the higher dimensional metric and discuss its properties. We show that, under precise conditions, this metric is always Ricci flat and admits a homothetic Killing vector field; if the evolving spacetimes are maximally symmetric, their ($d+1$)-dimensional representative has a vanishing Riemann tensor even. The non-degeneracy of the higher dimensional metric which ``geometrizes" a given RG trajectory is linked to a monotonicity requirement for the running of the cosmological constant, which we test in the case of Asymptotic Safety.
\end{abstract}
\thispagestyle{empty}

\newpage
\restoregeometry
\setcounter{page}{1}
\section{Introduction}
The familiar renormalization group (RG) equations of quantum field theory are formulated in a mathematical setting which is rather simple and, in a way, structureless from the geometric point of view. The only ingredients involved are a manifold $\mathscr{T}$, often referred  to as the theory space, and a vector field $\bm{\beta}$ thereon. The data $\left(\mathscr{T},\bm{\beta}\right)$ suffice to describe what is called the RG flow and to define the integral curves of $\bm{\beta}$ on $\mathscr{T}$. Since the components of $\bm{\beta}$ are given by the ordinary beta-functions, the first-order differential equations that govern these integral curves, aka RG trajectories, are nothing but the standard renormalization group equations. Even for the more general functional RG equations the situation is essentially the same, except for the infinite dimensionality of the manifold $\mathscr{T}$ whose points represent full fledged effective action functionals.

There is, however, a longstanding conjecture that beyond $\bm{\beta}$ further natural geometric objects might be ``living" on the manifold $\mathscr{T}$.
For example, after the advent of Zamolodchikov's c-theorem \cite{Zamolodchikov1, Zamolodchikov2}, related investigations in more than 2 dimensions focused on searching for a scalar ``c-function" and a metric on $\mathscr{T}$ by means of which the RG flow could be promoted to a \textit{gradient flow}. Even though this program was not fully successful in the generality originally hoped for, it ultimately led to important developments such as the proof of the a-theorem \cite{Komargodski}.

Furthermore, various authors, guided by different motivations have tried to furnish the manifold $\mathscr{T}$ with a \textit{connection} \cite{Lassig, Sonoda, Dolan}. Recently significant progress has been made along these lines after, in ref.\cite{Carlo1}, a powerful functional RG framework for the analysis of composite operators had been introduced. In this setting, the connection which has been proposed \cite{Carlo2} is related to the operator product expansion coefficients.

The conjectured \textit{AdS/CFT correspondence} ``geometrizes" RG flows by a different approach which identifies the scale variable of the RG equations with a specific coordinate on a higher dimensional (bulk) spacetime \cite{Maldacena, Gubser, Witten}. In this way the ``RG time" acquires a status similar to the ordinary spacetime coordinates.

Along a different line of research, the fundamental idea of \textit{dimensionally extending spacetime by scale variables} has been developed in considerable generality in the work of L. Nottale \cite{Nottale}. In his approach the RG time is on a par with the usual spacetime coordinates, both conceptually and geometrically.

\bigskip

\noindent \textbf{(1)} The present paper is devoted to a different notion of geometrized RG flows. While it does have certain traits in common with the various theoretical settings mentioned above, it is more conservative, however, in that its starting point does not involve any unproven assumptions. This starting point consists of nothing but the standard RG trajectories supplied by a functional renormalization group equation (FRGE). We propose to exploit those RG-derived data, and only those, to initiate a systematic search for \textit{natural geometric structures} which can help in efficiently structuring those data and/or facilitate their physical interpretation or application.\footnote{A first analysis along these lines can be found in \cite{Percacci1} where contact was made with the Randall-Sundrum model.}

\bigskip
\noindent \textbf{(2)} Specifically we deal in this paper with the nonperturbative functional RG flows of Quantum Einstein Gravity (QEG), i.e., quantum gravity in a metric-based formulation. We assume that it is described by an effective action functional $\Gamma_k[\cdot]$ which depends both on a 4D spacetime metric $g_{\mu \nu}$, and on some kind of RG scale, $k \in \mathbb{R}^+$, implemented as an infrared cutoff, for example. Furthermore we suppose that we managed to solve the corresponding FRGE for (partial) trajectories in theory space, i.e., maps $k \mapsto \Gamma_k[\cdot]$ whereby the curve parameter $k$ does not necessarily cover all scales $k \in \mathbb{R}^+$.

For every given value of $k$, the running effective action $\Gamma_k$ implies an effective field equation for the expectation value of the metric, typically a generalization of Einstein's equation. Solutions to those effective Einstein equations inherit a $k$-dependence from $\Gamma_k[\cdot]$, and we shall denote them $g^k_{\mu \nu}(x^\rho)$ in the following. More precisely, in this paper we are going to analyze a situation where the solutions at differing scales are selected such that $g_{\mu \nu}^k$ depends on $k$ smoothly. Therefore we may regard the map $k \mapsto g_{\mu \nu}^k(x^\rho)$ as a smooth trajectory in the space of all metrics that are compatible with a given differentiable manifold, $\mathscr{M}_4$. Thus, technically speaking, the output of the functional RG - and effective Einstein equations amounts to a family of Riemannian structures on one and the same spacetime manifold:
\begin{equation}
 \Big\{\big(\mathscr{M}_{4}, \;g_{\mu \nu}^k \big)\;\big| \;k \in \mathbb{R}^+\Big\}
	\end{equation}

\bigskip

\noindent \textbf{(3)} In this paper we propose a new way of thinking about the infinitely many metrics $g_{\mu \nu}^k$ that furnish the same 4D spacetime manifold $\mathscr{M_4}$. Namely, we shall interpret the family $\left(\mathscr{M}_{4}, g_{\mu \nu}^k \right)$ as different 4D slices through \textit{a single} 5-dimensional Riemannian or pseudo-Riemannian manifold:
\begin{equation}
\big(\mathscr{M}_{5}, \;{}^{(5)}g_{IJ} \big)
\end{equation}
Hereby all $g_{\mu \nu}^k$'s arise from only one 5D metric ${}^{(5)}g_{IJ}$ by isometrically embedding the slices into $\mathscr{M}_5$.

If $k$ has the interpretation of an (inverse) coarse graining scale on $\mathscr{M}_4$, then $\mathscr{M}_5$ naturally comes close to a ``scale-space-time" manifold \cite{Nottale}. In addition to the usual event coordinates $x^\mu$, its points involve a certain value of the scale or coarse graining parameter: $(k, x^\mu)$.

\bigskip
\noindent \textbf{(4)} Actually $\mathscr{M}_5$, equipped with some metric ${}^{(5)}g_{IJ}$ can encode more information than is contained in the underlying family $\left(\mathscr{M}_{4}, g_{\mu \nu}^k \right)$. This is most obvious if we use local coordinates which are adapted to the foliation by the surfaces of equal scale. The scale parameter (or an appropriate function thereof) plays the role of a $5^{\rm{th}}$ coordinate then, and the basic trajectory of 4D metrics $g_{\mu \nu}^k(x^\rho) \equiv g_{\mu \nu}(k, x^\rho)$ is reinterpreted as 10 out of the 15 independent components which ${}^{(5)}g_{IJ}(k, x^\rho)$ possesses.

Our main interest is in its additional components, ${}^{(5)}g_{\mu k}(k, x^\rho)$ and ${}^{(5)}g_{kk}(k, x^\rho)$, respectively. The question we are going to address is whether those functions can be determined in a mathematically or physically interesting way such that a single 5D geometry not only encapsulates or ``visualizes" a trajectory of 4D geometries, but also enriches it by additional information. Schematically,

\vs{-3}
\begin{equation}
	\tcboxmath{\text{trajectory of 4D geometries}}\;\; + \;\;\tcboxmath{ \; \text{?} \;}\quad= \quad\tcboxmath{ \text{\text{unique 5D geometry}}}
\end{equation}

\bigskip
\vs{2}
Loosely speaking, what we are proposing here is a bottom-up approach which starts out from the safe harbor of a well understood and fully general RG framework and only in a second step tries to assess whether, and under what conditions there exist natural options for geometrizing the RG flows.

This approach must be contrasted with top-down approaches like the one based upon the AdS/CFT conjecture, for instance. They would rather begin by postulating the geometrization, and ask about its relation to standard RG flows at the second stage only.

\bigskip

\noindent \textbf{(5)} The discussions in this paper are largely independent of the precise details concerning the underlying RG technique and the concrete trajectories $g_{\mu \nu}^k$. An exception is Section 6 below which makes essential use of the gravitational Effective Average Action \cite{Martin, Frank}. Implementing a \textit{Background Independent} \cite{Rovelli} coarse graining procedure in presence of dynamical gravity, it is ideally suited for the description of \textit{self-gravitating quantum systems} like the ones we shall consider \cite{Carlo3, Carlo4, Max}. While not restricted to this application \cite{Wetterich-2, Wetterich-1, Wetterich-0, Wetterich+1, Wetterich+2} the Effective Average Action has been used extensively in the Asymptotic Safety program \cite{Weinberg, Lauscher, Lauscher2, Lauscher3, Percacci, Percacci2, Frank}. There, its Background Independence is likely to be the essential ingredient responsible for the formation of a nontrivial RG fixed point \cite{Weyer}.

\bigskip

\noindent\textbf{(6)} The rest of this paper is organized as follows. In Section 2 we set up a convenient framework, based upon a generalization of the ADM construction, for the embedding of a given family $g_{\mu \nu}^k$ in a higher dimensional geometry. In Section 3 we then provide a simple, yet fully explicit and sufficiently general class of such families $g_{\mu \nu}^k$. They correspond to running Einstein metrics, and all subsequent demonstrations refer to this class of solutions.

A priori our goal of searching of ``interesting" 5D geometries is an extremely broad one; to be able to make practical progress we therefore narrow its scope to a particular class of ADM metrics, which we introduce and discuss in Section 4. Then, in Section 5, we derive our main results. We show that, under precise conditions, the running 4D Einstein spaces can always be embedded in a 5D geometry which admits a \textit{homothetic Killing vector field} and is \textit{Ricci flat}; should the Einstein spaces be maximally symmetric, it is even strictly, i.e., Riemann flat.

The important point about these options for a geometrization of RG flows is that they neither follow from pure geometry alone, nor are they ``for free" what concerns the properties of the RG trajectory. Rather, they are a global geometric manifestation of a specific general feature of the RG trajectory. In the present example, this \textit{sine qua non} is that the running cosmological constant $\Lambda(k)$ is a strictly increasing function of the scale. In Section 6 we show that for the asymptotically safe trajectories of QEG this is indeed the case. Finally, Section 7 contains a summary and the conclusions.

\section{From trajectories of metrics to higher dimensions}
	Let us suppose that we employed some sort of functional renormalization group (FRG) framework, whose specifics do not matter here, in order to derive a scale dependent effective field equation, i.e., a generalization of Einstein's equation. We assume that we furthermore managed to solve this one-parameter family of differential equations, thus obtaining families of metrics $g_{\mu \nu}^k$ labelled by the RG scale $k$.

	According to the standard interpretation outlined in \cite{Schwindt1, Schwindt2}, the set $\{g_{\mu \nu}^k\}_{k\geq0}$ gives rise to a \textit{family of different Riemannian structures} all of which furnish one and the same $d$-dimensional manifold $\mathscr{M}_d$. Correspondingly, one formally regards $k \mapsto (\mathscr{M}_d, g_{\mu \nu}^k)$ for $k \in \mathbb{R}^+$ as a ``trajectory" in the space of $d$-dimensional (Euclidean) spacetimes.
	
	In local coordinates we write their line elements as
\begin{equation}
\di s_d^2 = g_{\mu \nu}^k(x^\rho)\di x^\mu \di x^\nu, \qquad \mu, \nu , \cdots = 1,2,\cdots, d.
\end{equation}
	For generality we switched here from 4 to $d$ spacetime dimensions.
	\bigskip
	
\noindent \textbf{(1)} The key idea of the present work is to re-interpret the RG parameter $k$, possibly after a convenient reparametrization $\tau = \tau(k)$, as an additional \textit{coordinate} which, together with $x^\mu$, coordinatizes a ($d+1$)-dimensional manifold $\mathscr{M}_{d+1}$. The original manifold $\mathscr{M}_{d}$ is isometrically embedded in $\mathscr{M}_{d+1}$ in a $k$-dependent way, and so $\mathscr{M}_{d+1}$ comes into being equipped with a natural foliation.

According to this re-interpretation, the entire RG trajectory of ordinary spacetimes is described by \textit{a single} Riemannian structure on the higher dimensional manifold. We denote it $\left(\mathscr{M}_{d+1}, {}^{(d+1)} g_{IJ}\right)$, and write the corresponding line element as 
\begin{equation}
\di s_{d+1}^2 = {}^{(d+1)}g_{IJ}(y^K)\di y^I \di y^J
\end{equation}
where $y^I \equiv (y^0, y^\mu)$ are generic local coordinates on $\mathscr{M}_{d+1}$.

Here and in the following indices $I, J, K, \cdots$ always assume values in $\{0,1,2, \cdots, d\}$, while Greek indices run from 1 to $d$ only.

\bigskip

\noindent \textbf{(2)} Since $\mathscr{M}_{d+1}$ is endowed with a natural foliation it is convenient to employ a variant of the ADM formalism \cite{Frank}. To prepare the stage let us assume we are given an arbitrary manifold $\left(\mathscr{M}_{d+1}, {}^{(d+1)} g_{IJ}\right)$. We start by defining a scalar function $y \mapsto \tau(y)$ which assigns a specific scale to each of its points. Then we construct the level sets of this ``RG time function", $\Sigma_\tau \equiv \big\{y \in \mathscr{M}_{d+1}|\tau(y) = \tau\big\}$, and interpret them as a stack of standard, $d$-dimensional spacetimes that differ with respect to their resolution scale $\tau \equiv \tau(k)$, however.

The gradient $n_I \equiv N \partial_I \tau$ defines a vector $n^I$ which is everywhere normal to the slices $\Sigma_\tau$. By choosing the lapse function $N(y)$ appropriately, we normalize it such that
\begin{equation}
{}^{(d+1)}g_{IJ}\;n^I n^J = \varepsilon
\end{equation}
where $\varepsilon = \pm 1$ depends on whether the normal vector is space- or time-like.\footnote{We allow ${}^{(d+1)}g_{IJ}$ to be a Lorentzian metric of any signature. However, $g_{\mu \nu}^k$ is assumed to have Euclidean signature, unless stated otherwise.}

Next we transform from the generic coordinates $y^I = y^I(x^J)$ to new ones, $\displaystyle{x^I \equiv (x^0, x^\mu)}$ $\equiv (\tau, x^\mu)$, which are adapted to the foliation: $\tau$ labels different ``RG time slices" and the $x^\mu$'s are coordinates on a given $\Sigma_\tau$. Defining a vector field $t^I$ by the condition $t^I \partial_I\tau = 1$, we relate the coordinate systems on neighbouring slices by requiring that the coordinates $x^\mu$ are constant along the integral curves of $t^I$.

The tangent space at any point of $\mathscr{M}_{d+1}$ can be decomposed into a subspace spanned by vectors tangent to $\Sigma_\tau$ and its complement. The corresponding basis vectors are given by derivatives of the functions $y^I = y^I(x^J) = y^I (\tau, x^\mu)$ that describe the embedding of $\Sigma_\tau$ into $\mathscr{M}_{d+1}$:
\begin{equation}
e_\mu{}^I = \frac{\partial}{\partial x^\mu}y^I(\tau, x^\alpha), \qquad t^I = \frac{\partial}{\partial \tau}y^I(\tau, x^\alpha)
\label{def}
\end{equation}
As a result the $e_\mu$'s are orthogonal to $n$:
\begin{equation}
{}^{(d+1)}g_{IJ}\;n^I e_\mu{}^J = 0
\end{equation}
Furthermore, on the slices $\Sigma_\tau$, the embedding induces the following metric from the ambient metric ${}^{(d+1)}g_{IJ}$:
\begin{equation}
{}^{(d)}g_{IJ}= e_\mu{}^I e_\nu{}^J\;\;{}^{(d+1)}g_{IJ}
\end{equation}
In general the vector $t^I$ has nonvanishing components in the directions of both $n^I$ and $e_\mu{}^I$. Its expansion 
\begin{equation}
t^I = N n^I + N^\mu e_\mu{}^I
\end{equation}
involves the lapse function $N(\tau, x^\mu)$ and the shift vector $N^\mu (\tau, x^\mu)$. The definitions (\ref{def}) also entail that the coordinate one-forms in the two coordinate systems are related by 
\begin{equation}
\di y^I = t^I \di \tau + e_\mu{}^I \di x^\mu = N n^I \di \tau + e_\mu{}^I(\di x^\mu+ N^\mu \di \tau)
\label{this}
\end{equation}
Upon inserting (\ref{this}) into $ \di s_{d+1}^2=  {}^{(d+1)}g_{IJ}\di y^I \di y^J$ we obtain the line element recast in terms of the ADM variables $\{N, N^\mu, {}^{(d)}g_{\mu \nu}\}$:
\begin{equation}
\di s_{d+1}^2 = \varepsilon N(x^I)^2 \di \tau^2 + {}^{(d)}g_{\mu \nu}(x^I)\Big [\di x^\mu + N^\mu (x^I)\di \tau\Big]\Big[\di x^\nu + N^\nu (x^I)\di \tau\Big]
\label{ADM}
\end{equation}

The sign $\varepsilon = \pm 1$ which determines the signature of the higher dimensional metric is left open at this point. Later on we shall encounter criteria which determine whether the RG time $\tau$ really turns into a time coordinate ($\varepsilon = -1$) and describes a Lorentzian metric on $\mathscr{M}_{d+1}$, or whether it amounts to a further spatial dimension ($\varepsilon = +1$).

\bigskip

\noindent \textbf{(3)} To make contact with the RG approach, we assume that the higher dimensional metric has the ADM form (\ref{ADM}), and we then identify ${}^{(d)}g_{\mu \nu}$ with the output of the computations based upon the FRGE and the effective field equations:
\begin{equation}
{}^{(d)}g_{\mu \nu}(\tau, x^\rho) = g_{\mu \nu}^k (x^\rho)\Big|_{k = k(\tau)}
\label{metrics}
\end{equation}
The (invertible) function $k(\tau)$ amounts to an optional and physically irrelevant redefinition of the original scale parameter in terms of a convenient RG time $\tau$. A typical example is $\tau = \ln (k/\kappa)$, or even simpler, $\tau = k/\kappa$.\footnote{Unless stated otherwise all coordinates are dimensionless in our conventions, while all metric coefficients have mass dimension $-2$. Since $[k] = +1$, the constant $\kappa$ must have $[\kappa] = +1$.} In the following we shall assume that both $g_{\mu \nu}^k (x^\rho)$ and $k(\tau)$ are known, externally prescribed functions.

Thus, knowing ${}^{(d)}g_{\mu\nu}(x^I)$, what is still lacking in order to fully specify the higher dimensional line element (\ref{ADM}) are the lapse and shift functions $N$ and $N^\mu$, respectively, as well as the sign $\varepsilon$. These are properties of the metric on $\mathscr{M}_{d+1}$ which do \textit{not} follow from the flow equations.

\bigskip

\noindent \textbf{(4)} This leads us back to our main question: Is it conceivable that there exist general reasons or principles, over and above those inherent in the RG framework, that determine those missing ingredients in a meaningful and physically relevant way?

Inspired by the  familiar applications of the ADM formalism in General Relativity, one might be tempted to argue that there can be little physics in $N$ and $N^\mu$, since, by a $\diff(\mathscr{M}_{d+1})$ transformation, we can change them in an almost arbitrary way. It is important though to emphasize that \textit{this argument does not apply in the present context}.

The reason is that here the possibility of performing coordinate transformations has been exhausted already in solving the $k$-dependent effective field equations. The ADM framework in $d+1$ dimensions imports concrete functions $g_{\mu \nu}^k (x^\rho)$ from the RG side, and they refer to a specific set of coordinates. Since we do not want those functions to be changed by a $\diff(\mathscr{M}_{d+1})$ transformation, and we insist (for the time being) that they occupy the $\mu$-$\nu$ sub-matrix of ${}^{(d+1)}g_{IJ}$, we have to allow functions $N$ and $N^\mu$ of any form in this gauge picked by $g^k_{\mu \nu}$.

As a consequence, we first must arrive at a certain triple $\{N, N^\mu, \varepsilon\}$ which completes the specification of $\di s_{d+1}^2$, and only then we are free to perform general coordinate transformations, if we desire to do so.


\section{Solutions of rescaling type: running Einstein spaces}
To make the later discussion as explicit as possible, let us pause here for a moment and introduce a technically particular convenient class of running metrics whose $k$-dependence resides entirely in their conformal factor.

\bigskip

\noindent \textbf{(1)} To this end we assume that we are dealing with pure quantum gravity (no matter fields), and that the Einstein-Hilbert truncation is employed, meaning that the effective field equations are\footnote{In this paper we denote higher dimensional geometric objects (e.g. the curvature scalar ${}^{(d+1)}R$, etc.) by the prepended label ($d+1$), while all those without are the original ones referring to $\mathscr{M}_d$. In particular $R_{\mu \nu}$ denotes the Ricci tensor related to $g_{\mu \nu}$ in $d$ dimensions, while ${}^{(d+1)}R_{\mu\nu}$ are the $\mu$-$\nu$-components of the tensor ${}^{(d+1)}R_{IJ}$ built from ${}^{(d+1)}g_{IJ}$.}  $G_{\mu \nu}[g_{\alpha \beta}^k] = - \Lambda (k)g_{\mu \nu}^k$, or equivalently, with $R_\mu{}^\nu [g_{\alpha \beta}^k] = (g^k)^{\nu \rho} R_{\mu \rho}[g^k]$,
\begin{equation}
R_\mu{}^\nu [g_{\alpha \beta}^k] = \frac{2}{d-2}\;\Lambda(k)\;\delta_\mu{}^\nu
\label{fieldeq}
\end{equation}

\noindent \textbf{(2)} In this setting, the only input from the RG equations is the $k$-dependence of the running cosmological constant, $\Lambda(k)$. The latter can be of either sign, and it also might vanish at isolated scales. It will turn out convenient to express it in the form 
\begin{equation}
\Lambda (k) = \sigma\; |\Lambda(k)|
\nonumber
\end{equation}
with the piecewise constant sign function $\sigma = \pm 1$, and to introduce the quantity
\begin{equation}
H(k) = \left[\frac{2|\Lambda(k)|}{(d-1)(d-2)}\right]^{1/2}\label{hubble}
\end{equation}
in order to write the absolute value of the cosmological constant as
\begin{equation}
|\Lambda (k)| = \frac{1}{2} (d-1)(d-2)\;H(k)^2.
\end{equation}

For every fixed value of $k$, the solutions to the effective field equation
\begin{equation}
R_\mu{}^\nu[g_{\alpha \beta}^k] = \sigma (d-1)\;H(k)^2\; \delta_\mu{}^\nu
\label{Ricci}
\end{equation}
are arbitrary Einstein manifolds \cite{Besse} with scalar curvature 
\begin{equation}
	R[g_{\alpha \beta}^k] = \sigma \;d(d-1)\;H(k)^2. \label{Rh}
\end{equation}
Among them there are the distinguished ones which possess a maximum number of Killing vectors, namely the spheres $S^d$, pseudo-spheres $H^d$, and the flat space $R^d$. They exist for $\sigma = +1$ and $\sigma = -1$ when $H(k)\neq 0$, and for $H(k) = 0$, respectively.

The motivation for the $d$-dependent factors in the definition (\ref{hubble}) is as follows.
Comparing (\ref{Rh}) with the standard result for the curvature scalar of maximally symmetric spaces reveals that, for the special case when $g_{\mu \nu}^k$ is maximally symmetric, $1/H(k)$ is nothing but the radius of curvature of the corresponding sphere or pseudo-sphere. Thus $H(k)$ can be identified with the conventionally defined Hubble parameter. Hence the Riemann tensor is normalized as follows in the case of maximal symmetry:
\begin{equation}
R_{\mu \nu \rho \sigma}[g_{\alpha \beta}^k] = \sigma \;H(k)^2\;[g_{\mu \rho} g_{\nu \sigma} - g_{\mu \sigma} g_{\nu \rho}] \label{riemsym}
\end{equation}

We emphasize however that while we are going to employ the quantity $H(k)$ defined by (\ref{hubble}) as a convenient way of representing the cosmological constant, we are \textit{not} confining our attention to maximal symmetry in what follows.

\bigskip

\noindent\textbf{(3)} Coming back to the problem of finding solutions to (\ref{fieldeq}), let us fix some convenient reference scale $k_0$ at which
\begin{equation}
\Lambda(k_0) \equiv \Lambda_0 \equiv \frac{1}{2}\sigma \;(d-1)(d-2)\;H_0^2
\end{equation}
and let us pick an arbitrary solution $g_{\mu \nu}^{(0)}(x^\rho)$ of the classical vacuum Einstein equation involving this particular value of the cosmological constant:
\begin{equation}
R_\mu{}^\nu [g_{\alpha \beta}^{(0)}] = \sigma (d-1)\;H_0^2\;\delta_\mu{}^\nu
\label{einstein}
\end{equation}
It then follows that the ``running metric" given by
\begin{equation}
g_{\mu \nu}^k (x^\rho) = Y(k)^{-1} \;g_{\mu \nu}^{(0)}(x^\rho)
\label{Y}
\end{equation}
with $g_{\mu \nu}^k|_{k = k_0} = g_{\mu \nu}^{(0)}$, and 
\begin{equation}
Y(k) \equiv \frac{|\Lambda(k)|}{|\Lambda_0|} \equiv \frac{H(k)^2}{H_0^2}
\end{equation}
solves the effective field equation (\ref{fieldeq}) on all scales $k$ that are sufficiently close to $k_0$. This is to say that $\Lambda$ must not have any zero between $k$ and $k_0$ so that $\sigma = \text{sign}(\Lambda(k)) = \text{sign} (\Lambda_0)$ is a constant function. Equation (\ref{Y}) is easily proved by noting that the Ricci tensor, with mixed indices, behaves as 
\begin{equation}
R_\mu{}^\nu [c^{-2}g_{\alpha \beta}] =c^{2}R_\mu{}^\nu[g_{\alpha \beta}]
\end{equation}
under global Weyl transformations with an arbitrary real $c$.
\bigskip

\noindent \textbf{(4)} We now have a simple but, as we shall see, instructive example of a trajectory made of generic, i.e., not necessarily maximal symmetric, Einstein spaces at our disposal. Upon inserting (\ref{Y}) into (\ref{metrics}) the spacetime part of the ADM metric reads
\begin{equation}
{}^{(d)}g_{\mu \nu}(\tau, x^\rho)=Y(k(\tau))^{-1}\;g_{\mu \nu}^{(0)}(x^\rho)
\end{equation}
with the externally prescribed function $\tau \mapsto Y(k(\tau))$ coming from the RG machinery.

\section{Focusing on the lapse function}
Let us recall that it is our aim to explore the theoretical possibilities of fixing the missing ingredients of the higher dimensional metric, $\{N, N^\mu, \varepsilon\}$, in a way that is physically or mathematically distinguished, for one reason or another.

\bigskip

\noindent \textbf{(1)} As it stands, the scope of this investigation is extremely broad. In a first attempt it can help therefore to narrow down the setting in order to make the problem technically more clear-cut, and its physics interpretation more transparent.

For this reason we focus in the sequel on a vanishing shift vector, and on lapse functions that depend on $\tau$ only: $N^\mu (x^I) = 0$, $N(x^I) = N(\tau)$. As a consequence eq.(\ref{ADM}) boils down to
\begin{equation}
\di s_{d+1}^2 = \varepsilon \;N(\tau)^2\;\di \tau^2 + {}^{(d)}g_{\mu \nu} (\tau, x^\rho)\di x^\mu \di x^\nu
\end{equation}
As we shall see, this truncated form of the ADM metric is still sufficiently rich, and yet simple enough to allow for practical progress.
\bigskip

\noindent \textbf{(2)} Regarding the RG input, we now insert the explicit trajectory of Einstein metrics found in the previous section:
\begin{equation}
\di s_{d+1}^2 = \varepsilon \;N(\tau)^2\;\di \tau^2 +Y(k(\tau))^{-1}\; \;g^{(0)}_{\mu \nu} ( x^\rho)\di x^\mu \di x^\nu
\label{d+1}
\end{equation}
Thus, should there exist a yet to be discovered general principle that endows the metric on $\mathscr{M}_{d+1}$ with information that goes beyond the input data provided by the RG equations, this information must reside in the lapse function $N(\tau)$.

It is important at this point to remember that the $g^{(0)}_{\mu \nu}$'s are externally prescribed coefficient functions which we do not want to be changed by coordinate transformations. Hence, if some principle when applied to (\ref{d+1}) demands that the lapse must have a particular functional form $\tau \mapsto N(\tau)$, this lapse function refers to an already \textit{fully gauge fixed metric}, the corresponding gauge being selected in the process of solving the effective field equations.

\bigskip

\noindent\textbf{(3)} What is a single Riemannian or pseudo-Riemannian manifold $\left(\mathscr{M}_{d+1}, {}^{(d+1)}g_{IJ}\right)$ capable of doing for us that would not already be possible using the original stack of unrelated manifolds $\Big\{\left(\mathscr{M}_{d}, g_{\mu \nu}^k\right)\big|k \in \mathbb{R}^+\Big\}$?

One answer is that it can ascribe proper lengths to curves in $\mathscr{M}_{d+1}$ which are not confined to a single slice of the foliation. Such curves explore not only different points of spacetime, but also different scales.

As an example let us consider a curve $\mathcal{C}(P_1, P_2)$ connecting two points $P_{1,2} \in \mathscr{M}_{d+1}$. In the coordinate system of (\ref{d+1}), they are assumed to possess the same $x^\mu$-, but different $\tau$-coordinates, namely $\tau_1$ and $\tau_2$, respectively. The curve begins on the RG time slice having $k(\tau_1) = k_1$ and ends on the one with $k(\tau_2) = k_2$. Furthermore, we assume that $x^\mu = \text{const}\equiv c^\mu$ is constant along $\mathcal{C}(P_1, P_2)$, so that the curve projects onto a single point of $\mathscr{M}_d$.

Then, for $\varepsilon = +1$ say, the metric in eq.(\ref{d+1}) tells us that this curve has the proper length
\begin{equation}
\Delta s_{d+1} \equiv \int_{\mathcal{C}_{(P_1,P_2)}} \sqrt{\di s_{d+1}^2} \;\;=\;\; \int_{\tau_1}^{\tau_2}\di \tau N(\tau)
\label{deltas}
\end{equation}
Loosely speaking, this integral allows us now to answer questions like: ``What is the distance between a high-scale object and a low-scale object at one and the same spacetime event ($x^\mu$)?"

In more realistic examples $P_1$ and $P_2$ may have different $x^\mu$-coordinates so that $\mathcal{C}$ visits more than one point of $\mathscr{M}_d$. Hence the two ``objects" need not to lie on top of one another. The resulting proper length $\Delta s_{d+1}$ is a mixture then of the familiar distance in spacetime, and the separation of the two objects in the scale direction.

If $\Delta s_{d+1}$ is to have any physical meaning it must be possible to experimentally connect coordinates ($\tau, x^\mu$) to the results of certain measurements. A well known model for achieving this on ordinary spacetimes equips $\mathscr{M}_{d}$ with a set of scalar fields $\phi^\mu$ whose observable values represent $x^\mu$ then \cite{Ferrero}. In the case at hand we must invoke an additional field which allows a determination of the scale $k (\tau)$. In cosmology, say, one might think of a local temperature field, for instance.

\bigskip

\noindent \textbf{(4)} As we mentioned earlier, the function $k(\tau)$ can be chosen freely. It is gratifying to see therefore that the propertime $\Delta s_{d+1}$ in (\ref{deltas}) is indeed independent of this choice. Assume that two such functions belong to the same foliation, i.e., $k(\tau) = k = \bar k (\bar \tau)$, and the respective RG times are related by the coordinate transformation $\tau = \tau (\bar \tau)$. The latter belongs to the foliation-preserving subgroup of $\diff(\mathscr{M}_{d+1})$, and it acts on the lapse function according to \cite{Frank}
\begin{equation}
\bar N (\bar \tau) = N(\tau)\left(\frac{\di \tau}{\di \bar \tau}\right)
\end{equation}
As a consequence, $\bar N(\bar \tau) \di \bar \tau = N(\tau) \di \tau$, and (\ref{deltas}) is seen to be invariant.

\section{Distinguished higher dimensional geometries}

The crucial question is what kind of physical or mathematical principle could possibly determine the higher dimensional metric, and what are the universal geometric features of the manifold $\left(\mathscr{M}_{d+1}, {}^{(d+1)}g_{IJ}\right)$ which result from it. The information coming from the RG trajectory determines ${}^{(d+1)}g_{IJ}$ only incompletely. Using the prescribed coordinate system of eq.(\ref{d+1}), what is left to be determined by this principle are $N(\tau)$ and $\varepsilon$.
\vs{10}

\bigskip
In this paper, we postulate that the RG trajectories under consideration possess the following monotonicity property:
\begin{flalign}
	\qquad\textbf{(P) } \text{The cosmological constant } \Lambda (k) \text{ is a strictly increasing function of } k.&&
\end{flalign}
Taking \textbf{(P)} for granted, we are going to demonstrate that it is always possible to complete the specification of the $(d+1)$-dimensional (pseudo-) Riemannian geometry in such a way that it enjoys the following features:
\begin{flalign}
	\qquad \textbf{(G) } \text{The higher dimensional metric } {}^{(d+1)}g_{IJ} \text{ is } \textit{Ricci flat}\text{: } {}^{(d+1)}R_{IJ} = 0.&&
\end{flalign}
This property is universal in the sense that it pertains to arbitrary $k$-dependent Einstein metrics $g_{\mu \nu}^k(x^\rho)$.

Furthermore, if the $d$-dimensional Einstein metrics $g_{\mu \nu}^k(x^\rho)$ happen to be \textit{maximally symmetric}, but still curved in general, then \textbf{(G)} can be replaced by the stronger statement:
\begin{flalign}
\qquad \textbf{(G\textquotesingle) } \text{The higher dimensional metric } {}^{(d+1)}g_{IJ} \text{ is}\textit{ Riemann flat}\text{: } {}^{(d+1)}R^I{}_{JKL} = 0.&&
\end{flalign}

In this section we are going to show that \textbf{(G)} and \textbf{(G\textquotesingle)}, respectively, are indeed made possible by \textbf{(P)} since it allows us to postulate a highly distinguished and universal form of ${}^{(d+1)}g_{IJ}$. Thereafter we shall investigate whether the postulated property \textbf{(P)} is actually realized in Asymptotic Safety.

\subsection{The Hubble length as a coordinate}
We consider $k$-intervals with different signs of $\Lambda (k)$ separately, should they occur. If $\Lambda(k)>0$, \textbf{(P)} entails that $Y(k)$ and $H(k)$ are monotonically increasing with the scale, while the Hubble length
\begin{equation}
L_H(k) \equiv \frac{1}{H(k)} = \left[\frac{(d-1)(d-2)}{2|\Lambda(k)|}\right]^{1/2}
\label{hlength}
\end{equation}
is a decreasing function of $k$. If instead $\Lambda(k)<0$, the postulate \textbf{(P)} requires $Y(k)$ and $H(k)$ to decrease, and $L_H(k)$ to increase with $k$. In either case the postulated strict monotonicity implies that the function $L_H(k)$ is invertible, i.e., the relationship between $k$ and $L_H$ is one-to-one. A a consequence, we may regard the map $k \mapsto L_H(k)$ given by (\ref{hlength}) as a reparametrization of the ``scale manifold" $\mathbb{R}^+$ or a subset thereof, and $L_H$ as a concrete example of an RG time $\tau = \tau (k)$. Up to now, the $\tau$-$k$ relationship has been an arbitrary convention. Here now we make a specific choice for this coordinate, not by hand but by invoking the RG trajectory itself. 

For clarity we denote this special RG time coordinate by $\xi$. The corresponding coordinate transformation $k = k(\xi)$ is determined by the implicit condition
\begin{equation}
\xi \equiv L_H\big(k (\xi)\big),
\label{LH}
\end{equation}
while its inverse is known explicitly:
\begin{equation}
\xi(k) = \left[\frac{(d-1)(d-2)}{2|\Lambda(k)|}\right]^{1/2}
\label{xi}
\end{equation}

When we reexpress $\di s^2_{d+1}$ in terms of $\xi$ we are led to the conformal factor
\begin{equation}
Y(k(\xi))^{-1} = H_0^2\;\; H(k(\xi))^{-2} = H_0^2\;\;L_H(k(\xi))^2 = H_0^2\; \xi ^2
\end{equation}
Hence, in the new system of coordinates, the second term on the RHS of eq.(\ref{d+1}) assumes a very simple dependence on the RG time, being proportional to $\xi^2$.


The sought-for principle that decides about the $(d+1)$-dimensional geometry, after having installed the coordinates\footnote{The special RG time $\xi$ and its ``conformal" analogue $\eta$ to be introduced below are the only exceptions to our rule that coordinates are dimensionless. While $[x^\mu] = 0 $ throughout, $\xi$ and $\eta$ have the dimension of a \textit{length}: $[\xi] = [\eta] = -1$.} $x^I \equiv (x^0, x^\mu) \equiv(\xi , x^\mu)$, must come up with a unique function $\xi \mapsto N(\xi)$. This, then will allow us to completely specify the line element $\di s_{d+1}^2 \equiv {}^{(d+1)}g_{IJ} (x^K)\di x^I \di x^J$ in eq.(\ref{d+1}).
\bigskip

In order to prove in a constructive way that the postulate \textbf{(P)} indeed allows us to achieve \textbf{(G)} or \textbf{(G\textquotesingle)}, respectively, we enact the following rule for the completion of ${}^{(d+1)}g_{IJ}$:
\begin{flalign}
	\quad \;\;\textbf{(R) } \textit{In the } (\xi, x^\mu) \textit{ system, } \text{the lapse function must assume the simplest form possible,}	&&
	\nonumber
\end{flalign}
\vs{-11}
\begin{flalign}
\qquad\;\;\;\;\; \text{     namely } N(\xi) = 1.&&
\end{flalign}
In other words, the coordinates realizing the ``proper RG time gauge" are required to coincide with those which employ the Hubble length as the scale coordinate. The rule \textbf{(R)} enforces that the higher dimensional metric is unambiguously given by
\begin{equation}
\di s_{d+1}^2 = \varepsilon\; (\di \xi)^2 + \xi^2\;\;H_0^2\; \;g_{\mu \nu}^{(0)} (x^\rho) \di x^\mu \di x^\nu
\label{complete}
\end{equation}
which is fully determined except for the $\text{sign}\;\varepsilon$.

We emphasize that the property \textbf{(P)} is crucial for making the rule \textbf{(R)} meaningful. Without the strict monotonicity of $\Lambda(k)$ we could not have replaced $k$ with $\xi \propto |\Lambda(k)|^{-1/2}$ in its role as a coordinate.\footnote{For a similar discussion of coordinate transformations on the $g$-$\lambda$ theory space see ref.\cite{Percacci}.}

Note that the metric (\ref{complete}) possesses a remarkable \textit{universality property}: It has no explicit dependence on the function $\Lambda(k)$. In the $(\xi, x^\mu)$ coordinate system, the proposed metric ``remembers" $\Lambda(k)$ only via the implicit requirement that $\xi \leftrightarrow |\Lambda(k)|^{-1/2}$ must be one-to-one.

In the $(\xi, x^\mu)$ system, the information about the actual RG evolution resides entirely in the ``time function" $k = k(x^I) \equiv k(\xi, x^\mu)$ which describes how the slices $\Sigma_\tau \equiv \Sigma_\xi$ are embedded into $\mathscr{M}_{d+1}$. In the case at hand the time function has no dependence on $x^\mu$, and boils down to $k = k(\xi)$. It is this function that has been adjusted in (\ref{LH}) by imposing $\xi = L_H (k (\xi))$. Since the inverse function $\xi = \xi(k)$ is given by (\ref{xi}), we recover the time function belonging to the line element (\ref{complete}) by solving $\xi \propto |\Lambda(k)|^{-1/2}$ for $k = k(\xi)$.

Eq.(\ref{complete}) is our proposal for the single higher dimensional metric which ``geometrizes" the entire RG history of the original metrics.\footnote{It is interesting to note that the metric (\ref{complete}) plays a prominent role also in the 5D ``space-time-matter theory" in \cite{Wesson1, Wesson2, Wesson3}.}  In the following subsections we are going to discuss its detailed properties which, as a matter of fact, are the actual motivation for this specific proposal.

 \subsection{Equivalent forms of the postulated metric}
 The special status of the ($\xi, x^\mu$) system of coordinates resides solely in the fact that in this system the lapse function is defined to be particularly simple, namely $N = 1$. After having set up the metric ${}^{(d+1)}g_{IJ}$ we may transform it to any coordinate system $\bar x^I \equiv (\bar x^0, \bar x^\mu)$ we like. Here we mention two simple foliation preserving transformations.
 
 \bigskip
 
\noindent \textbf{(1) The conformal RG time.} In practical computations it is often convenient to transfer the scale dependence from the conformal factor of $g^{(0)}_{\mu \nu}$ to the overall conformal factor of ${}^{(d+1)}g_{IJ}$. This is achieved by the ($x^\mu$-independent!) transformation trading $\xi \in \mathbb{R}^+$ for $\eta \in \mathbb{R}$ via $\xi = H_0^{-1}e^{H_0 \eta}$, or conversely,
\begin{equation}
\eta = H_0^{-1}\; \ln (H_0 \xi) = L_H^0\; \ln (\xi/L_H^0),
\end{equation}
 with $L_H^0 \equiv H_0^{-1} \equiv  L_H (k_0)$. The new coordinate $\eta$ is positive (negative) if the length $\xi$ is of super- (sub-) Hubble size, according to the metric at the reference scale $k_0$. In the ($x^0 \equiv \eta, x^\mu$) system the line element (\ref{complete}) assumes the desired form:
 \begin{equation}
\di s_{d+1}^2 = e^{2 H_0\eta} \;\Big [\varepsilon\; (\di \eta)^2 + g_{\mu \nu}^{(0)} (x^\rho) \di x^\mu \di x^\nu\Big]
\label{eta}
 \end{equation}
 While, in its original form (\ref{complete}), $\xi$ is reminiscent of the cosmological time in a Robertson-Walker metric, the new variable $\eta$ has the interpretation of the corresponding conformal RG time.
 \bigskip
 
 \noindent \textbf{(2) The IR cutoff as a coordinate.} Both in the ($\xi, x^\mu$) and the ($\eta, x^\mu$) system of coordinates the metric is independent of $\Lambda(k)$, while the time functions $k= k(\xi)$ and $k = k(\eta)$ know about it. We can reverse the situation and make the time function trivial by introducing directly the cutoff $k$ (or the dimensionless $L_H ^0 \;k$) as the new coordinate. The change of coordinates $\xi \rightarrow k$ defined by (\ref{xi}) brings the metric (\ref{complete}) to the form
 \vs{2}
  \begin{equation}
	\di s_{d+1}^2 =\left|\frac{\Lambda_0}{\Lambda(k)}\right|\;\Bigg \{\varepsilon\; \left(\frac{1}{2} \partial_k \ln |\Lambda(k)|\right)^2 \; \left(L_H^0\; \di k\right)^2+ g_{\mu \nu}^{(0)} (x^\rho) \di x^\mu \di x^\nu\Bigg\}\label{IRcoord} \vs{2}
\end{equation}
 which is manifestly $\Lambda(k)$-dependent.
 
 We see that the metric (\ref{IRcoord}) degenerates at points where $\partial_k \Lambda(k) = 0$, hinting at the importance of \textbf{(P)} again. Note also that the proposed metric ascribes a nonzero distance to high- and low-scale objects at the same $x^\mu$ only when there is a non-trivial RG running, $\partial_k \Lambda(k) \neq 0$, so that the effective spacetimes acquire fractal properties \cite{Lauscher1, Frank2}.
 
 \subsection{Homothetic Killing vector and self-similarity}
 The ($d+1$)-dimensional geometry described by equation (\ref{complete}), or equivalently by (\ref{eta}), is a very particular one in that it admits a homothetic Killing vector field $X \equiv X^I \partial_I$. With $\mathscr{L}_X$ denoting the Lie derivative along $X$, this vector field satisfies the defining equation 
  \begin{equation}
\mathscr{L}_X \;\;{}^{(d+1)}g_{IJ} = 2\: C \;\;{}^{(d+1)}g_{IJ}
\label{killing}
 \end{equation}
 for $C = H_0$. Note that (\ref{killing}) differs from the condition for a generic conformal Killing vector field since $C$ is a constant rather than an arbitrary function on $\mathscr{M}_{d+1}$ \cite{Hall}.
 
 The homothetic vector field is explicitly given by
 \begin{equation}
X = \frac{\partial}{\partial \eta} = H_0 \;\xi \frac{\partial}{\partial \xi}
 \end{equation}
 It is easily checked therefore that it generates $x^\mu$- independent rescalings of the metric. The existence of such a vector field is the hallmark of self-similarity in the general relativistic context \cite{Carr}. It is a coordinate independent manifestation of the underlying foliation with self-similar leaves which may be hidden if inappropriate coordinates are used.
 
 \subsection{Ricci flatness}
 Finally we turn to the curvature of the postulated higher dimensional geometry. In order to better appreciate its rather unique character, we consider the following slightly more general class of metrics:
  \begin{equation}
{}^{(d+1)}g_{IJ}(x^K)\di x^I \di x^J = \Omega^2 (\eta)\Big [\varepsilon (\di \eta)^2 + g_{\mu \nu}^{(0)} (x^\rho) \di x^\mu \di x^\nu\Big]
\label{conformal}
 \end{equation}
 Here we employ the same coordinates $x^K \equiv (x^0 = \eta, x^\mu)$ as in eq.(\ref{eta}), but we admit for a moment an arbitrary overall conformal factor $\Omega (\eta)$.
 
 Working out the Ricci tensor of (\ref{conformal}) one finds\footnote{Our curvature conventions are $R^\sigma{}_{\rho \mu \nu} = + \partial_\mu \Gamma^\sigma_{\nu \rho}- \cdots$ and $R_{\mu \nu} = R^\rho{}_{\mu \rho \nu}$.}: 
  \begin{subequations}
 	\begin{align}
 		{}^{(d+1)}R^0{}_0 & = -\varepsilon \;d \;\Omega^{-2}\left[\frac{\ddot{\Omega}}{\Omega}- \left(\frac{\dot \Omega}{\Omega}\right)^2\right]\label{r00} \\
 		{}^{(d+1)}R^0{}_\mu & = 0, \qquad 	{}^{(d+1)}R^\mu{}_0  = 0\\
 		{}^{(d+1)}R^\mu{}_\nu & = \Omega^{-2}\left\{ R^\mu{}_\nu - \varepsilon\; \delta^\mu{}_\nu\left[\frac{\ddot{\Omega}}{\Omega}+(d-2) \left(\frac{\dot \Omega}{\Omega}\right)^2\right] \right\} \label{rmix}
 	\end{align}
 \end{subequations}
Here $R^\mu{}_\nu$ denotes the Ricci tensor of $g_{\mu \nu}^{(0)}(x^\rho)$, and the dot indicates derivatives with respect to $\eta$.

\bigskip

Now let us ask under what circumstances (\ref{conformal}) is Ricci flat:
\begin{equation}
{}^{(d+1)}R^I{}_J = 0
\end{equation}
By (\ref{r00}), the necessary and sufficient condition for ${}^{(d+1)}R^0{}_0 =0$ is that $\Omega\; \ddot{\Omega} = (\dot \Omega)^2$. The most general solution to this differential equation is given by
\begin{equation}
\Omega(\eta) = e^{B(\eta -\eta_0)}
\label{Omega}
\end{equation}
with arbitrary real constants $B$ and $\eta_0$. Using this form of $\Omega$ in (\ref{rmix}), the condition $	{}^{(d+1)}R^\mu{}_\nu =0$ is found to be equivalent to
\begin{equation}
R^\mu{}_\nu - \varepsilon(d-1)\;B^2\; \delta^\mu{}_\nu = 0
\label{cond2}
	\end{equation}
	
	Up to this point, $g^{(0)}_{\mu \nu}(x^\rho)$ and so the Ricci tensor $R^\mu{}_\nu$ have been left unspecified. When we now exploit that  $g^{(0)}_{\mu \nu}(x^\rho)$ actually describes an Einstein space complying with eq.(\ref{einstein}), the condition (\ref{cond2}) boils down to $\sigma H_0^2 - \varepsilon B^2 = 0$. This latter equation has the unique solution $\varepsilon= \sigma$, $B = H_0$.
	
	Thus the conclusion is that there does exist a \textit{Ricci flat} higher dimensional metric of the form (\ref{conformal}) for every $d$-dimensional metric with $g^{(0)}_{\mu \nu}$ describing a (curved) Einstein space. Furthermore, this metric is essentially unique and is obtained by letting
	\begin{equation}
\varepsilon = \sigma \qquad \text{and} \qquad \Omega (\eta) = e^{H_0(\eta-\eta_0)} \label{conditions}
	\end{equation}
in the family of line elements (\ref{conformal}):
	\begin{equation}
\di s^2_{d+1} = e^{2H_0(\eta-\eta_0)}\;\Big[\sigma \;(\di \eta)^2+ g_{\mu \nu}^{(0)}(x^\rho)\di x^\mu \di x^\nu\Big]
\end{equation}

Choosing $\eta_0 = 0$ brings is back to the metric (\ref{eta}) which we set out to investigate, with an additional piece of information, however. Originally we had admitted an arbitrary sign $\varepsilon = \pm 1$. But now we see that Ricci flatness can be achieved only if we allow the sign of the cosmological constant $\sigma = \Lambda_0/|\Lambda_0|$ to determine the signature of ${}^{(d+1)}g_{IJ}$. If the cosmological constant is positive (negative) the scale parameter becomes a spacelike (timelike) coordinate.

\subsection{Strict flatness}
Let us go one step further and ask under what conditions metrics of the form (\ref{conformal}) are not only Ricci flat but even strictly, i.e., Riemann flat:
\begin{equation}
{}^{(d+1)} R^I{}_{JKL} = 0.
\end{equation}
Modulo the usual symmetries, the Riemann tensor of (\ref{conformal}) has only the following nonzero components:
  \begin{subequations}
	\begin{align}
		{}^{(d+1)}R^{0\mu}{}_{0\nu} & = \varepsilon\;  \Omega^{-2}\left[ \left(\frac{\dot \Omega}{\Omega}\right)^2 - \frac{\ddot{\Omega}}{\Omega}\right]\delta^\mu{}_\nu \label{riem00}\\
		{}^{(d+1)}R^{\mu\nu}{}_{\rho \sigma} & = \Omega^{-2} \left\{R^{\mu \nu}{}_{\rho\sigma}-\varepsilon\left(\frac{\dot \Omega}{\Omega}\right)^2\Big[\delta^\mu{}_\rho\delta^\nu{}_\sigma- \delta^\mu{}_\sigma\delta^\nu{}_\rho\Big]\right\} \label{riemmix}
	\end{align}
\end{subequations}
Herein $R^{\mu\nu}{}_{\rho \sigma} $ is the Riemann tensor that belongs to $g_{\mu \nu}^{(0)}(x^\rho)$.

\bigskip

Imposing $	{}^{(d+1)}R^{0\mu}{}_{0\nu} =0$, eq.(\ref{riem00}) reproduces the requirement $\Omega\; \ddot{\Omega} = (\dot \Omega)^2$ and (\ref{Omega}) as its general solution. Inserting this solution into (\ref{riemmix}), the vanishing of the second set of components, $	{}^{(d+1)}R^{\mu\nu}{}_{\rho \sigma} = 0$, implies the following condition on the curvature tensor of $g^{(0)}_{\mu \nu}(x^\rho)$:
\begin{equation}
	R^{\mu\nu}{}_{\rho \sigma} = \varepsilon B^2\Big[\delta^\mu{}_\rho\delta^\nu{}_\sigma- \delta^\mu{}_\sigma\delta^\nu{}_\rho\Big]\label{riemann}
\end{equation}
The tensor structure on the RHS of (\ref{riemann}) is the hallmark of a maximally symmetric manifold, see eq.(\ref{riemsym}). We conclude therefore that the metric (\ref{conformal}) is strictly flat if and only if, \textit{first}, $\Omega(\eta)$ and $\varepsilon$ are fixed according to (\ref{conditions}), and \textit{second}, the running Einstein metric at $k = k_0$, i.e., $g_{\mu \nu}^{(0)}$, corresponds to a maximally symmetric $d$-dimensional space.

\bigskip

This completes our demonstration that, under this symmetry constraint, the geometric feature \textbf{(G)} of the higher dimensional manifold can be tightened to \textbf{(G\textquotesingle)}.

It is indeed quite remarkable that the inclusion of the scale variable has ``flattened" the curved spacetime. In the case at hand, the metric (\ref{complete}) specializes to
\begin{equation}
 \qquad\di s_{d+1}^2 =(\di \xi)^2+\xi^2 \;\di \Omega_d^2 \qquad\quad (\Lambda_0 > 0) \label{sphere}
\end{equation}
\begin{equation}
 \qquad\di s_{d+1}^2 =-(\di \xi)^2+\xi^2 \;\di H_d^2 \qquad (\Lambda_0 < 0) \label{hyperbola}
\end{equation}
where $\di \Omega_d^2$ and $\di H^2_d$ are the line elements for, respectively, $S^d$ and $H^d$ with unit length scale. Both of these metrics are well known to be flat: Equation (\ref{sphere}) describes ($d+1$)-dimensional Euclidean space in spherical coordinates. Hereby $\xi$ plays the role of the radial variable, $\mathscr{M}_{d+1} \equiv R^{d+1}$ being foliated by $d$-spheres of radius $\xi$.

Similarly the metric (\ref{hyperbola}) describes Minkowski space $M^{1,d}$. The RG time has become a genuine time coordinate in this case. Here Minkowski space is foliated by hyperbolic $d$-spaces whose radius of curvature is given by $\xi$. For $d=3$, eq.(\ref{hyperbola}) is nothing but the metric of Milne's universe.
 
 \section{Asymptotic Safety}
 In the previous section we saw that the ``principle" or ``property" \textbf{(P)} is a necessary condition for being able to define ${}^{(d+1)}g_{IJ}$ as a (Ricci) flat metric in the higher dimensional sense. In the present section we are going to discuss the actual situation concerning the monotonicity of $\Lambda(k)$ within the concrete setting of pure quantum gravity (QEG) in $d=4$ dimensions.
 
 We employ the prototypical Einstein-Hilbert truncation of the Effective Average Action, the first one used to demonstrate Asymptotic Safety \cite{Martin, Lauscher, Frank1}. The truncation is based upon the ansatz 
 \begin{equation}
 	\Gamma_k = \frac{1}{16 \pi \; G(k)} \int \di^4 x \;\sqrt{g}\; \Big(-R(g) + 2\Lambda(k)\Big) + \cdots
 	\label{EH}
 	\end{equation}
 where the dots indicate the classical gauge fixing and ghost terms. The resulting RG equations for the running couplings $G(k)$ and $\Lambda(k)$ were obtained in \cite{Martin} and solved numerically in \cite{Frank1}. In the following we are particularly interested in the properties of the function $\Lambda (k)$ along typical RG trajectories.
 \bigskip
 
 \noindent \textbf{(1) Mode counting functions.}  It is quite remarkable that considerations about 5D representations of the histories of 4D geometries have led us to scrutinize the monotonicity properties of $\Lambda(k)$. In fact, in ref. \cite{Becker} a closely related question, the monotonicity of the dimensionless product $G(k)\;\Lambda(k)$, has been explored already, for an entirely different reason though.
 
 In \cite{Becker} a c-function-like quantity $\mathscr{C}(k)$ has been proposed in 4D quantum gravity which, when evaluated exactly, should be monotonically decreasing along RG trajectories, and be stationary at fixed points. In simple truncations $\mathscr{C}(k)$ is proportional to $\big[G(k)\;\Lambda(k)\big]^{-1}$. Not unlike Zamolodchikov's c-function, $\mathscr{C}(k)$ can be argued to count the number of the fluctuation modes already integrated out, thus explaining its monotonicity when evaluated exactly. As for approximate calculations, it was found however that the above Einstein-Hilbert truncation is \textit{not} precise enough to render $\mathscr{C}(k)$ monotonic, while it does turn out monotone if we use more general truncations \cite{Becker1} of the bi-metric type \cite{Manrique, Manrique1, Manrique2}.
 
 It is not unreasonable to expect that $\Lambda(k)$ might have similar, if not better, mode counting properties. After all, in the most naive picture every bosonic fluctuation mode which is not suppressed by the cutoff contributes a positive zero-point energy to the cosmological constant and therefore should contribute additively to $\Lambda(k)$.
 \bigskip

 \noindent\textbf{(2) The trajectories simplified.} The classification of the RG trajectories implied by the ansatz (\ref{EH}) on the $g$-$\lambda-$plane of the dimensionless Newton constant $g$ and cosmological constant $\lambda$ is well known \cite{Frank1}. Here we focus on the three main classes, i.e., trajectories of Type Ia, Type IIa, and Type IIIa, respectively.
 
 \bigskip
 
 \noindent \textbf{(i)} All of these trajectories approach a non-Gaussian fixed point $(g_*, \lambda_*)$ when $k \to \infty$. In particular the dimensionless cosmological constant behaves as $\lambda(k) \equiv \Lambda(k)/k^2 \to \lambda_*$ in the asymptotic region. Hence
 \begin{equation}
\qquad \qquad  \Lambda (k) = \lambda_* \;k^2 \qquad\qquad(k \gtrsim \hat k)
 \end{equation}
 is a reliable approximation to the exact trajectory in this regime. It extends from ``$k = \infty$" down to a scale $\hat k$ which is of the order of the Planck mass $m_{\rm{Pl}} \equiv G_0^{-1/2}$ typically.
 
 \bigskip
  \noindent \textbf{(ii)} Below a relatively complicated, but short transition regime near $\hat k$, all trajectories of the above three types enter a semiclassical regime within which the behaviour of $\Lambda(k)$ is easy to describe again. At least qualitatively, the following simple formula provides a reliable approximation:
  \begin{equation}
\Lambda(k) = \Lambda_0 + \nu \;G_0\; k^4
  \end{equation}
 Here $\nu > 0$ is a scheme dependent constant, and the infrared values $\Lambda_0 \equiv \Lambda (k = 0)$ and $G_0 \equiv G(k = 0)$ arise as constants of integration whose values select a specific RG trajectory in the 2D theory space. The three types of trajectories differ with respect to the value of $\Lambda_0$. We have $\Lambda_0 < 0$, $\Lambda_0 = 0$, and $\Lambda_0 > 0$ for trajectories of Type Ia, IIa, and IIIa, respectively.

 If $\Lambda_0 \neq 0$ it is convenient to introduce the two length scales
 \begin{equation}
 \ell \equiv \left(\frac{\nu \;G_0}{|\Lambda_0|}\right)^{1/4}\quad, \qquad  L \equiv \left(\frac{\lambda_*}{|\Lambda_0|}\right)^{1/2}
 \end{equation}
Hence, in the semiclassical regime, 
 \begin{equation}
\Lambda(k) = |\Lambda_0 |\;\;\Big(\ell^4 \;k^4 \pm 1\Big)
\end{equation}
where the plus sign (minus sign) applies to the Type IIIa (Type Ia).
\bigskip

\noindent \textbf{(iii)} When $\Lambda_0 \neq 0$ the following ``caricature" of the function $\Lambda(k)$ is useful:
 \begin{equation}
\Lambda(k) = |\Lambda_0| \; \cdot\;\;\left\{ \begin{array}{ll}
	\ell	^4\;k^4 \pm 1 \qquad \text{  for } \quad 0\leq k \lesssim\hat k
	\\L^2 \; k^2 \qquad \quad \quad\text{for }\quad \quad k \gtrsim \hat k
\end{array}  
	\right.\ 
	\label{IandIII}
\end{equation}
It should be a reliable approximation, except possibly during a short interval of scales near $\hat k$ where the transition between the two regimes takes place. We shall investigate this transition regime separately below.

In the case $\Lambda_0 = 0$, the corresponding approximation reads instead
 \begin{equation}
	\Lambda(k) = |\Lambda_0|\; \cdot\;\;\left\{ \begin{array}{ll}
	\nu\;m_{\rm{Pl}}^{-2}\; k^4 \qquad \text{  for } \quad 0\leq k \lesssim\hat k
		\\\lambda_*\;k^2 \qquad \quad \quad\text{for }\quad \quad k \gtrsim \hat k
	\end{array}  
	\right.\ 
	\label{IIa}
\end{equation}
Eq.(\ref{IIa}) applies to the single trajectory of Type IIa, the separatrix \cite{Frank1}.

\bigskip
\noindent \textbf{(iv)} Regarding the monotonicity, we observe that, whenever (\ref{IandIII}) and (\ref{IIa}) are applicable, the dimensionful cosmological constant $\Lambda(k)$ is indeed a strictly monotonic function of $k$, and all trajectories of the Types Ia, IIa, and IIIa have the crucial property \textbf{(P)}.

\bigskip
\noindent \textbf{(3) The signature.} A second important piece of information concerning $\Lambda(k)$ is the piecewise constant sign function $\sigma (k) \equiv \Lambda(k)/|\Lambda(k)|$. Eqs.(\ref{IandIII}) and (\ref{IIa}) yield 
 \begin{equation}
\quad	\;\text{for Type Ia: }\qquad\;\;\varepsilon (k) =\;\;\left\{ \begin{array}{ll}
		-1 \qquad \text{ for } \quad 0\leq\ k <\ell^{-1}
		\\
		+1 \qquad\text{ for }\quad \quad k>\ell^{-1}
	\end{array}  
	\right.\quad
\end{equation}
 \begin{equation}
	\text{for Type IIa: }\qquad\;\varepsilon (k) =\;\;+1\qquad\text{ for all }\quad \quad k\geq0\qquad\;
\end{equation}
 \begin{equation}
	\text{for Type IIIa: }\qquad\varepsilon (k) =\;\;+1\qquad\text{ for all }\quad \quad k\geq0\qquad\;
\end{equation}

Thus we conclude that everywhere along RG trajectories of the Types IIa and IIIa the RG ``time" amounts to a \textit{spatial coordinate} actually. Starting out from a Euclidean spacetime $\mathscr{M}_4$ with signature $(+++\:+)$, the proposed geometrization of the RG flow leads us unavoidably to a manifold $\mathscr{M}_5$ having $(+++++)$.
\bigskip

For trajectories of the Type Ia the situation is more complicated. They display an intermediate scale $k = \ell^{-1}$ at which the cosmological constant vanishes, $\Lambda(\ell^{-1})=0$. When $k$ passes this special scale, the solutions to the effective field equations undergo a change of topology. Coming from above, the scalar curvature changes from $R[g_{\alpha \beta}^k]>0$, via $R[g_{\alpha \beta}^{1/\ell}] = 0$, to $R[g_{\alpha \beta}^k]<0$. In the maximally symmetric case, for example, this topology change corresponds to a sequence of spaces $S^4 \to R^4 \to H^4$.

We note however that at the present stage of its development Asymptotic Safety cannot yet describe topology change processes in a dynamical fashion, neither in physical time nor in RG time. For this reason we adopt a conservative attitude here and consider the two branches of the Ia trajectories, having $\Lambda(k) > 0$ and $\Lambda(k)<0$, respectively, as two unrelated (incomplete) trajectories and, at this stage, study them separately.

The upper branch ($k > \ell^{-1}$) of a Type Ia trajectory augments the Euclidean $\mathscr{M}_4$ to an, again, Euclidean $\mathscr{M}_5$ having signature $(+++++)$, while its lower branch ($k<\ell^{-1}$) gives rise to a Lorentzian 5D manifold with $(-++++)$.

It is quite intriguing to speculate that an RG trajectory of this kind could underlie a mechanisms of \textit{chronogenesis}: the emergence of time in an a priori purely Euclidean system.

\begin{figure}[H]
	\centering
	\includegraphics[scale=0.54]{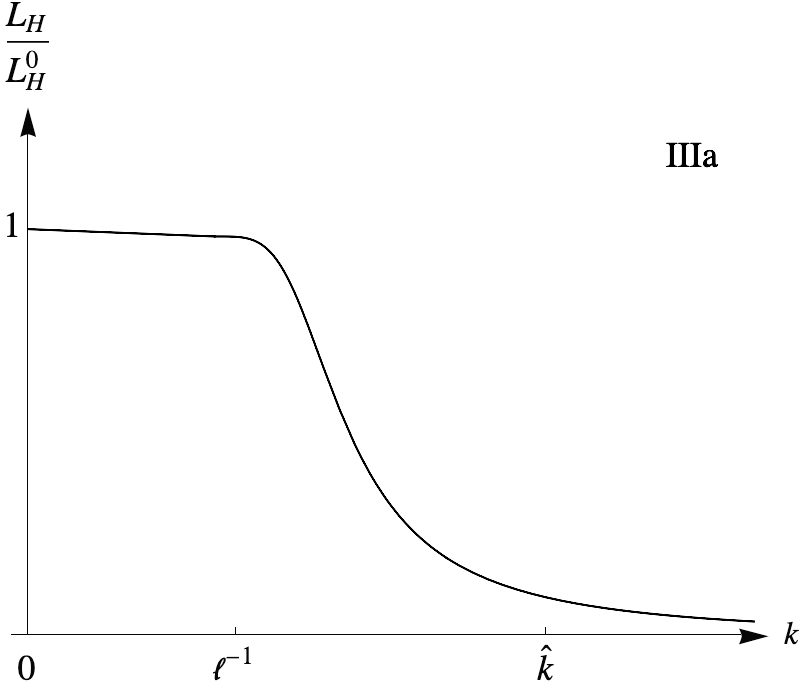}
	\quad
	\includegraphics[scale=0.54]{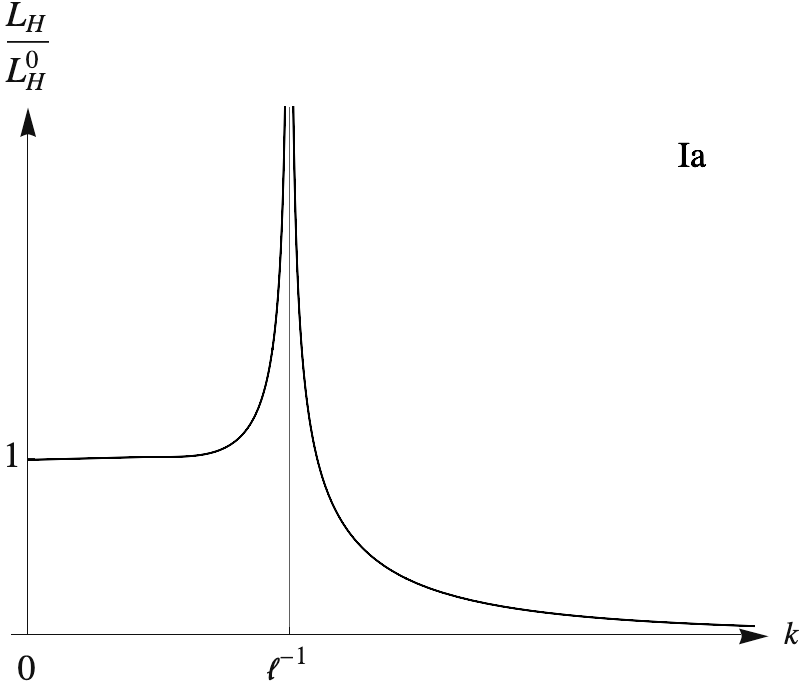}
	\caption{The scale dependent Hubble length along trajectories of Type IIIa (left) and Type Ia (right), respectively.}\label{Hubble}
\end{figure}

\bigskip
\noindent \textbf{(4) The coordinate change.} The dimensionless function $Y(k)$ can be written as
\begin{equation}
Y(k) = \frac{|\Lambda(k)|}{|\Lambda_0|} = \frac{H(k)^2}{H_0^2} = \left(\frac{L_H^0}{L_H(k)}\right)^2
\end{equation}
with $\Lambda_0 = 3H_0^2$ in $d=4$, and $L_H^0 \equiv 1/H_0$. Hence eq.(\ref{IandIII}) yields the following ``running Hubble length" $L_H(k) = L_H^0 Y(k)^{-1/2}$ along the trajectories of Type IIIa (plus sign) and of Type Ia (minus sign), respectively:
 \begin{equation}
L_H(k) = L_H^0\; \cdot\;\;\left\{ \begin{array}{ll}
\displaystyle	\frac{1}{\sqrt{|\ell^4 \; k^4 \pm 1|}}\qquad \text{  for } \quad 0\leq k \lesssim\hat k
		\\\displaystyle\frac{1}{L\;k} \qquad \qquad \qquad\text{ for }\quad \quad k \gtrsim \hat k
	\end{array}  
	\right.\ 
\end{equation}
This function is sketched in Figure \ref{Hubble}.

For the Type IIIa trajectories, the Hubble length $L_H(k)$ is seen to be a strictly decreasing function of the scale for all $k \in (0, \infty)$. It is one-to-one therefore, and so $\xi = L_H(k)$ defines a legitimate change of coordinates on the interval $(0, \infty)$. The same is true in the limiting case $\Lambda_0\searrow0$, i.e., for the Type IIa case.
\bigskip

\vs{-1.5}
Type Ia trajectories on the other hand decompose into two branches with $k \in(0, \ell^{-1})$ and $k \in(\ell^{-1}, \infty)$, respectively. On each branch separately, setting $\xi = L_H(k)$ is an allowed change of coordinates. On the upper (lower) branch the RG time $\xi$ becomes a strictly decreasing (increasing) function of $k$ then. However, employing $\xi$ globally would create a 2-1 ambiguity where $L_H > L_H^0$.

\bigskip
\noindent \textbf{(5) The transition region.} Finally let us investigate more carefully the monotonicity question in the transition region near $\hat k = O(m_{\rm{Pl}})$. The RG flow linearized about the fixed point ($g_*, \lambda_*$) is useful for a first orientation here. The linearization is governed by a pair of complex conjugate critical exponents $\theta_{1,2} = \theta' \pm i \theta''$, with $\theta'$, $\theta'' \in \mathbb{R^+}$, which are responsible for the spiral shaped trajectories $k \mapsto \left(g(k), \lambda(k)\right)$ encircling the fixed point. The latter is located in the first quadrant of the $g$-$\lambda-$ plane: $g_*>0$, $\lambda_*>0$. In the linear regime the condition $\partial_k\Lambda(k)>0$, or equivalently $k \partial_k \lambda(k) + 2 \lambda(k)>0$, assumes the form
\begin{equation}
\lambda_*+\zeta\; \left(\frac{k_0}{k}\right)^{\theta'}\; \cos \Big(\theta''\; \ln (k/k_0 )+ \alpha\Big) >0
\label{monoton}
\end{equation}
Here $\alpha$ and $\zeta$ are dimensionless parameters which depend on the constants of integration, that is, on the trajectory under consideration, as well as on the eigenvectors of the stability matrix.\footnote{See eq.(5.30) of ref.\cite{Lauscher}.} Eq.(\ref{monoton}) shows that for $k$ sufficiently large the monotonicity condition can never be violated since the potentially negative cosine is multiplied by too small a coefficient to compete with the positive $\lambda_*$. On the other hand, once the scale is low enough for $\zeta (k_0/k)^{\theta'}$ to be of order unity, there exist parameters $\alpha$, $\zeta$ for which (\ref{monoton}) could be violated. However, at those low scales the linear approximation is not necessarily valid any longer.
If by then the trajectory is already in the semiclassical regime, the ``caricature" trajectory applies and monotonicity is guaranteed; but if not, violations could occur.

A detailed numerical analysis reveals however that in reality there are no such violations of monotonicity in the transition region. For all three types of trajectories one finds that $\partial_k \Lambda(k) >0$ on all scales. 
	\begin{figure}[H]
	\centering
	\includegraphics[scale=0.56]{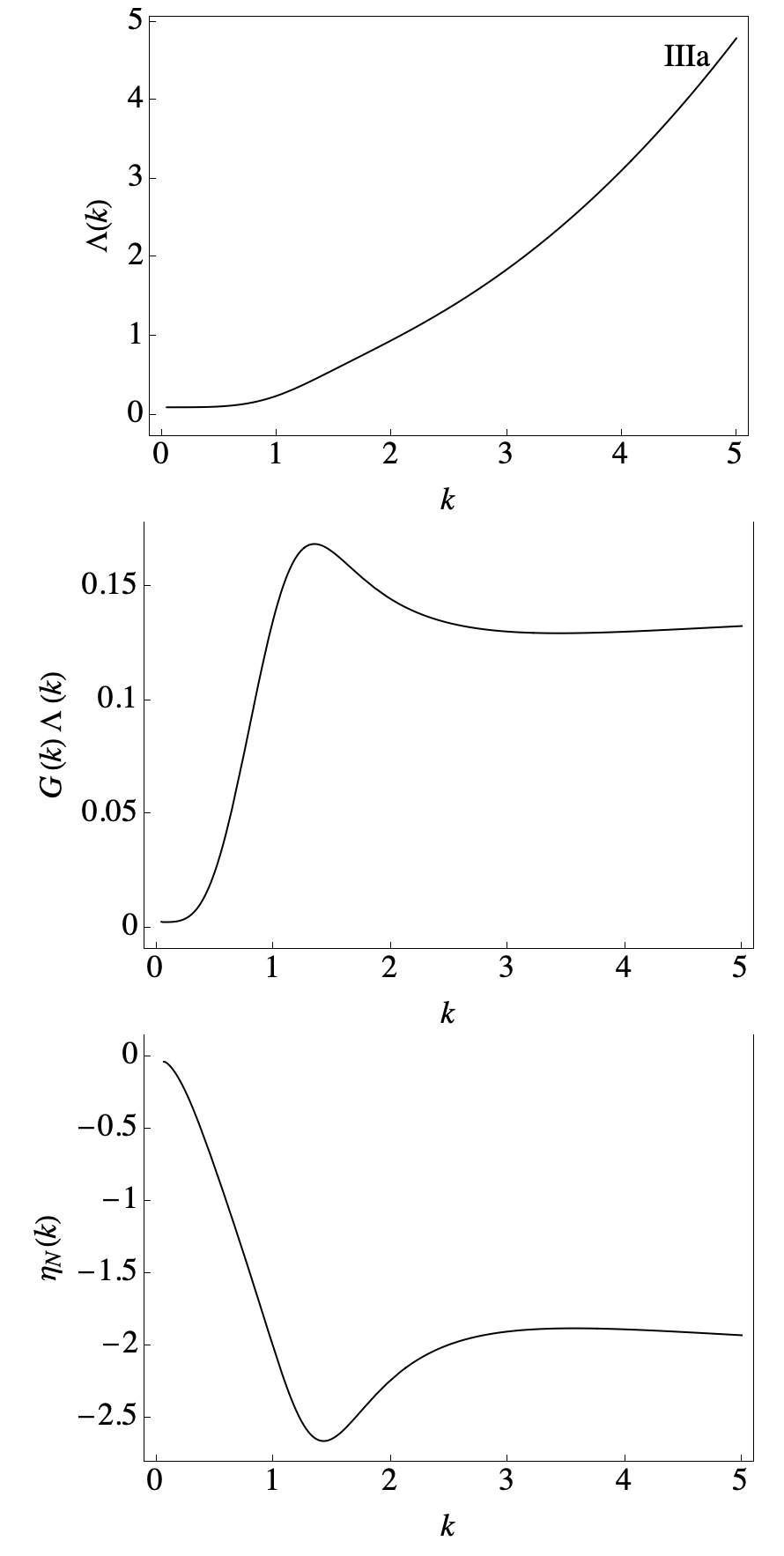}
	\caption{Numerical results for $\Lambda(k)$, the product $G(k)\; \Lambda(k)$, and the anomalous dimension $\eta_N(k)$ along a typical Type IIIa trajectory.}\label{Mono}
\end{figure}
Figure \ref{Mono} displays the numerical result for $\Lambda (k)$ and compares it to the product $G(k)\; \Lambda(k)$ and the anomalous dimension of Newton's constant, $\eta_N(k)$, along the same trajectory. The example shown is of Type IIIa, but since the plot focuses on the transition region it would look basically the same for the other types. It is quite impressive to see that $\Lambda(k)$ is indeed perfectly monotonic even in the transition regime, while this is by no means the case for $G(k)\;\Lambda(k)$ and $\eta_N(k)$. In particular the anomalous dimension displays significant oscillations in the transition regime.
\bigskip

This completes our demonstration that the asymptotically safe trajectories of QEG in 4 dimensions do indeed comply with the general property \textbf{(P)} and are thus eligible for a geometrization based upon the proposed rule \textbf{(R)}.

\section{Summary and Conclusion}
In this paper, we advocated a bottom-up approach towards geometrizing, and thus subsuming and visualizing the entire family of all effective spacetime metrics which occur along a given RG trajectory supplied by the established apparatus of the functional renormalization group for gravity. Different members of this family correspond to coarse grained 4D spacetimes at different resolutions.

The proposed geometrization is constituted by a single 5D Riemannian or pseudo-Riemannian manifold, $\mathscr{M}_5$. It carries a natural foliation whose leaves are the 4D spacetimes corresponding to a fixed RG scale. The RG trajectory which delivers the ``input data" for this construction determines the geometry of $\mathscr{M}_5$ only incompletely. A single metric on $\mathscr{M}_5$ can encode more information than the collection of all metrics on the slices. This raises the question if there exist any distinguished ways of completing the specification of the 5D geometry. Such completions might, for instance, be ``natural" from the mathematical perspective, or they could transport additional physics information that is not, or not easily accessible by the RG methods. 

It was one of the motivations for this paper to initiate a survey of the logical possibilities concerning such distinguished higher dimensional geometries which is unbiased with regard to particular geometries or models (AdS, Randall-Sundrum, etc.). Nevertheless, a long-term goal of this search program is to ultimately try making contact with ``top-down" formalisms like the AdS/CFT approach which also invoke scale-space-times, but bear no obvious relation to the effective average action and its functional RG flows.

\bigskip
As a proof of principle we explicitly analyzed the simplified situation where the ADM metric on $\mathscr{M}_5$ has a vanishing shift vector; we also assumed that the RG evolution of the 4D metrics is purely multiplicative, and that it is governed by the Einstein-Hilbert truncation of the effective average action. Under these conditions we proved that it is always possible to complete the specification of the 5D geometry in such a way that it possesses the following distinctive features: First, the metric on $\mathscr{M}_5$ admits a \textit{homothetic Killing vector field} as an intrinsic characterization of its self-similarity, and second, the metric on $\mathscr{M}_5$ is \textit{Ricci flat}. In the special case of maximally symmetric 4D spacetimes it even can be chosen \textit{strictly flat}.

These results are based upon a specific proposal for the general structure of the full 5D metric, eq.(\ref{complete}). Surprisingly enough, in the literature this class of metrics had already been studied in considerable detail, for quite different reasons though, namely in connection with the ``space-time-matter theory" advocated in \cite{Wesson1, Wesson2, Wesson3}.

From a more physics oriented point of view it is remarkable that in order to be well defined, i.e., non-degenerate, the proposed metric requires the cosmological constant $\Lambda(k)$ to be a \textit{strictly increasing} function of the cutoff $k$. With other words, the coefficient $\Lambda(k)$ in the effective average action must have properties similar to a c-function that ``counts" the number of fluctuation modes which get integrated out when $k$ is changed.
\bigskip

It is intriguing therefore to speculate that ultimately the envisaged geometrization encodes global information about the underlying flows that is not easily seen at the FRGE level. Hence future work will have to focus on drawing a more complete picture by relaxing some, or perhaps all of our assumptions.


\end{document}